\documentclass[conference]{IEEEtran}

\makeatletter
\def\ps@headings{%
\def\@oddhead{\mbox{}\scriptsize\rightmark \hfil \thepage}%
\def\@evenhead{\scriptsize\thepage \hfil \leftmark\mbox{}}%
\def\@oddfoot{}%
\def\@evenfoot{}}
\makeatother
\usepackage{url}
\usepackage{graphicx, subfigure}
\usepackage{enumerate}
\usepackage{balance}
\usepackage[usenames,dvipsnames]{color}
\usepackage{multirow}
\usepackage{array}
\usepackage{verbatim}
\usepackage{amsmath, amsthm, amssymb}

\newcommand{\vct}[1]{\ensuremath{\mathbf{#1}}}

\newcommand{\phishdef}{{\tt PhishDef}}
\newcommand{\ie}{{\em i.e., }}
\newcommand{\eg}{{\em e.g., }}

\begin{document}

% Title Page
\title{PhishDef: URL Names Say It All}

\author{Anh Le, Athina Markopoulou\\
University of California, Irvine\\
{\tt \{anh.le, athina\}@uci.edu}
\and
Michalis Faloutsos \\
University of California, Riverside\\
{\tt michalis@cs.ucr.edu}
}

\date{}

\maketitle

%--------------- Abstract -----------------------------------------------------
\begin{abstract}
Phishing is an increasingly sophisticated method to steal personal user information using sites that pretend to be legitimate. In this paper, we take the following steps to identify phishing URLs.  First, we carefully select lexical features of the URLs that are resistant to obfuscation techniques used by attackers. Second, we evaluate the classification accuracy when using only lexical features, both automatically and hand-selected, vs. when using additional features. We show that lexical features are sufficient for all practical purposes. Third, we thoroughly compare several classification algorithms, and we propose to use an online method (AROW) that is able to overcome noisy training data. Based on the insights gained from our analysis, we propose \phishdef, a phishing detection system that uses only URL names and combines the above three elements. \phishdef~is a highly accurate method (when compared to state-of-the-art approaches over real datasets), lightweight (thus appropriate for online and client-side deployment), proactive (based on online classification rather than blacklists), and resilient to training data inaccuracies (thus enabling the use of large noisy training data).

\begin{comment}
Phishing is an emerging and increasingly sophisticated method to steal personal user information using sites that pretend to be legitimate. Various detection methods exist but they are usually (a) reactive, by relying on the use of URL blacklists, (b) heavyweight, by checking an extensive list of properties, such as network properties, or (c) sensitive to the accuracy of training data. To overcome these weaknesses, we propose an approach for identifying phishing URLs on-the-fly using only lexical features. First, we carefully select lexical features of the URLs that are resistant to the obfuscation techniques that hackers often use. Second, we evaluate the classification accuracy when using only lexical features (both automatically and hand-selected) vs. when using additional features. Third, we propose to use an online classification method (AROW) that is able to overcome noisy training data.

Our key insights from our analysis of real datasets are that (a) lexical features are sufficient for identifying phishing URLs, (b) our selection of lexical features can detect even obfuscated URLs, and (c) our approach is highly resilient to noise. Based on the above observations, we propose \phishdef, a phishing detection system that is highly accurate (compared to state-of-the-art approaches over real datasets); lightweight (thus appropriate for online and client-side deployment); proactive (based on online classification rather than blacklists); and resilient to training data inaccuracies (thus enabling the use of large noisy training data.)
\end{comment}

\end{abstract}

%--------------- Introduction --------------------------------------------------
\section{Introduction}
\label{sec:intro}
Phishing is continuously evolving and becoming an increasingly sophisticated criminal tool to steal sensitive information and commit crimes on the Internet. According to the latest report from the Anti-Phishing Working Group \cite{AntiPhish}, the number of commercial brands being attacked by phishing just hit a new record: 356 brands in October 2009. With major industry targets, such as, financial and payment services, phishing has caused billions of dollars loss annually \cite{Gartner}. Because of the severity of the problem, the Internet community has put a significant amount of effort into defense mechanisms. 

Currently, two of the most popular services that protect the Internet users from visiting phishing sites are the Google Safe Browsing service \cite{GSB} and the Microsoft Smart Screen service \cite{MSS}. Both services provide client browsers with URL blacklists. The browsers, in turn, protect users from visiting the blacklisted URLs. The major problem of this protection model is that it is reactive: a phishing URL can only be included in the blacklist if it has already appeared somewhere else, {\em e.g.}, in a spam email, or has been reported by a user. A proactive model, where brand new phishing URLs could be identified accurately, is highly desirable to better protect the users. 

We argue that in order to provide a proactive protection, the machine learning classification engine, which is typically used to maintain the blacklists at the server  side, must be pushed to the client browser. This would allow new URLs to be classified on-the-fly, at the time the users click on or type in the URLs. One of the biggest challenges of classifying URLs on-the-fly, as opposed to off-line at the server side, is the latency constraint. The longer it takes to obtain the classification result of a URL, the longer a user has to wait to load that URL, and the worse the user experience. Furthermore, since page loading time is a decisive factor when benchmarking web browsers, classifying URLs should not introduce high latency.

There are two types features that can be used in URL classification: {\em lexical features}, {\em i.e.}, features which are readily available from the URL names; and {\em external features}, {\em i.e.}, features acquired from queries to remote servers. We refer to lexical and external features together as {\em full features}. Lexical features are based only on the URL names and are appropriate for implementation at the client. External features rely on the availability of remote servers, introduce additional latency due to the required queries, and consume more resources of the client, {\em e.g.}, battery life and bandwidth of mobile phones. Nonetheless, one would expect that relying on a more comprehensive set of features, rather than lexical features only, would lead to higher classification accuracy. In this paper, we seek to answer the following question:

\begin{quote} How well can one detect phishing URLs using only lexical features compared to using full features? \end{quote}

{\flushleft To} the best of our knowledge, this work is the first to extensively study this question. We show that lexical features are sufficient (\ie if properly used, they can achieve accuracy comparable to full features), and we propose a system called \phishdef~that achieves this goal.

In particular, we first introduce a way to extract lexical features that are resistant to obfuscation. We then thoroughly evaluate the classification accuracy achieved when using lexical features vs. full features with several state-of-the-art learning algorithms on real datasets. More specifically, we consider the following algorithms: batch-based Support Vector Machine (SVM), Online Perceptron (OP), Confidence-Weighted (CW), and Adaptive Regularization of Weights (AROW). We find that, using lexical features results in a modest decrease (about 1\%) in classification accuracy compared to using full features; however, the overall accuracy is still high (96--98\%). This suggests that using lexical features is sufficient and provides a better latency-accuracy trade-off. Moreover, our proposed obfuscation-resistant lexical features help to boost the overall classification accuracy across all the datasets. In particular, the reduction of error rate is up to 27\%. We also observe that state-of-the-art online linear classification algorithms, namely, AROW and CW, are more accurate while imposing less memory and computing overhead compared to other techniques. Moreover, when there is noise in the training data (noisy labels), AROW outperforms CW. Robustness in a noisy environment is very important because (i) it allows for training more comprehensive classification models by working with larger datasets, which typically include noise, such as, blacklisted URLs from Google used in \cite{GooglePhish}; and (ii) it improves the system's resilience to poisoning attacks, where attackers attempt to maliciously influence the classification models by injecting mis-labeled data.

Based on the insights gained from our analysis, we propose \phishdef, a classification engine that operates at the client side, uses only lexical features, and implements the AROW algorithm. \phishdef~has the following desired properties:
\begin{itemize}
\item High accuracy: It has 96--97\% classification accuracy, only 1\% less than full features.
\item Light-weight: It has low latency and imposes a modest amount of memory and computation overhead.
\item Proactive approach: It can classify new URLs on-the-fly, {\em i.e.}, at the time the user clicks on or enters the URL at the client side, as opposed to reactively relying on blacklists.
\item Resilience to noise: It maintains high accuracy even when trained with mislabeled data: 95\%--86\% accuracy when there is 5\%--45\% noise.
\end{itemize}

The rest of this paper is organized as follows. Section  \ref{sec:related} discusses related work. Section \ref{sec:dataset} describes the datasets we use and the feature extraction process. Section \ref{sec:classification} describes the classification algorithms we compare. Section \ref{sec:evaluation} presents the evaluation results, \ie the comparison of all algorithms over all datasets and feature sets. Section \ref{sec:analysis} discusses and explains the classification performance. Section \ref{sec:deployment} presents \phishdef,  our proposed solution based on the insights from the analysis. Section \ref{sec:conclusion} concludes the paper.

%--------------- Related Work ------------------------------------------------
\section{Background}
\label{sec:related}
%\subsection{Definition of Phishing}

PhishTank \cite{PhishTank} defines phishing as ``a fraudulent attempt to get you to provide personal information, including but not limited to, account information.'' This definition is somewhat restricted. In this work, we adopt a broader definition of phishing from Whittaker {\em et al.} \cite{GooglePhish}, which defines a phishing page as ``any web page that, without permission, alleges to act on behalf of a third party with the intention of confusing viewers into performing an action with which the viewer would only trust a true agent of the third party.''\footnote{This definition covers the typical case of phishing pages -- pages that mimic financial companies' sites and request login credentials from the viewers -- and also phishing pages that display trusted companies' logos to trick the viewers to download and execute malicious binary.}

%\subsection{Related Work}
Garera {\em et al.} \cite{GareraWorm} studied the structure of phishing URLs. They find four distinct categories of obfuscation techniques that phishing URLs use. Based on these categories, they propose eighteen manually selected features that can help to produce high classification accuracy. Their selected features include both lexical features and external features, such as Google PageRank and Google page quality of the page. Part of our work builds on these identified categories. We also propose features that address the four common obfuscation techniques, which, however, are directly extractable from the URL strings.

 Whittaker {\em et al.} \cite{GooglePhish} describes the design of the Google's phishing classifier used to automatically maintain Google's phishing blacklist. This classifier uses a wide variety of features: from lexical features, such as whether the URL contains an IP address, to URL metadata, such as Google PageRank, as well as features extracted from the page content and hosting information. While this work describes the classifier used to maintain blacklists at the server side, our work focuses on the design of an on-the-fly classifier at the client side.

In \cite{MaKDD}, Ma {\em et al.} examine the performance of several batch-based learning algorithms on classifying malicious URLs, which include phishing URLs and URLs present in spam emails. The algorithms are evaluated when working with various feature sets, for instance, host-based features, such as, features from WHOIS queries, and lexical features. This work shows that the combination of host-based and lexical features results in the highest classification accuracy. This work also hinted that using lexical features may lead to high accuracy; however, it did not investigate this direction in sufficient depth. Our work builds on this initial observation. We extensively evaluate how both batch-based and online algorithms perform when using only lexical features compared to full features.

In a follow-up work, Ma {\em et al.} \cite{MaICML} compare the performance of batch-based algorithms to online algorithms when using full features. The authors find that online algorithms, especially Confidence-Weighted (CW), outperform batch-based algorithms. Our main difference from \cite{MaICML} is that we focus on lexical features instead of full features. We propose obfuscation-resistant lexical features, show that online algorithms outperform batch-based algorithms when working with lexical features, and provide detailed analysis of the datasets to explain why this is the case. In addition, we introduce the use of AROW, which performs as well as CW but outperforms CW when there is noise. To the best of our knowledge, AROW has not been used before in the phishing context.

Other related work include PhishNet \cite{PhishNet}, which proposes heuristics to predict phishing URLs; the comparative analysis of phishing and non-phishing URLs drawn from PhishTank \cite{PhishTank} and DMOZ \cite{DMOZ} by McGrath and Gupta \cite{McGrath}; CANTINA \cite{CANTINA}, which uses a weighted sum of 8 features (4 content-related, 3 lexical, and 1 WHOIS) to classify phishing URLs; the classification of phishing emails by Fette {\em et al.} \cite{Fette} and Bergholz {\em et al.} \cite{Bergholz}; and the comparison of various tools for detecting fake websites by Abbasi and Chen \cite{Abbasi}.

Besides Google Safe Browsing \cite{GSB} and Microsoft Smart Screen \cite{MSS}, mentioned in the introduction, there are other commercial products which aim at protecting users from phishing sites, such as, McAfee SiteAdvisor \cite{McAfee} and WOT Web of Trust \cite{WOT}. The former incorporates proprietary feature analysis, and the latter relies on community feedbacks. These approaches are based on blacklists, thus reactive.

%--------------- Data Collection and Feature Extraction ----------------------------------------
\section{Datasets and Feature Extraction}
\label{sec:dataset}

%-------------------------- Dataset
\subsection{Malicious and Legitimate URLs}

{\flushleft \bf PhishTank.} PhishTank \cite{PhishTank} is a community site where anyone can submit, verify, and share phishing URLs. A suspected phishing URL will be manually checked by at least 2 other members of the site. Once verified as a phishing URL, the URL will be included in a downloadable database. We collect our set of phishing URLs during the one month period of June 2010.  The set consists of 4,082 verified phishing URLs ordered by their submission time.

{\flushleft \bf MalwarePatrol.} 
MalwarePatrol \cite{MalwarePatrol} is a free and user contributed system where anyone can submit suspicious URLs that may carry malware, viruses, or trojans. If a submitted URL is verified as malicious by MalwarePatrol, the URL will be put into a downloadable blacklist. We collect 2,001 malicious URLs during the last two weeks of June 2010. We order these URLs by their appearance time. We note that the URLs here have different characteristics from the URLs from PhishTank because they are crafted to spread malware while the URLs from PhishTank are crafted to steal sensitive information.

{\flushleft \bf Yahoo Directory.} Our first set of benign URLs is collected from the Yahoo directory. Yahoo provides a generator URL \cite{YahooDir}, which randomly generates a URL in its directory whenever someone visits it. We used this generator URL in mid June 2010 to collect 4143 random URLs. 

{\flushleft \bf Open Directory.} We collect our second set of benign URLs from DMOZ \cite{DMOZ}, which is one of the largest open directory of the Web maintained by volunteer editors. We collect 4012 random URLs from DMOZ directory in mid June 2010. 

For the benign URLs, we order them by the order in which we obtain them. We also note that our methodology of collecting URL datasets is similar to recent work \cite{MaKDD,PhishNet}.

%-------------------------- External Features
\subsection{External Feature Collection}
\label{subsec:external}

\begin{figure}[t]
\vspace*{-5pt}
\centering
\includegraphics[width=6cm]{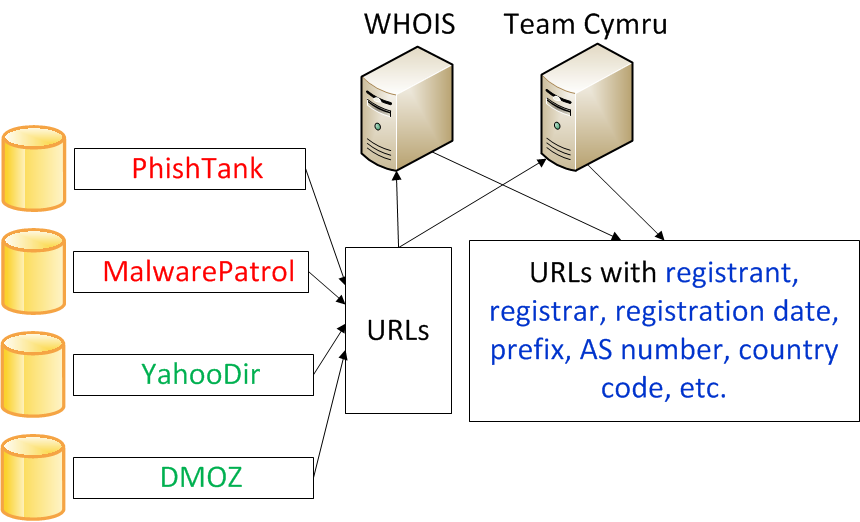}
\vspace*{-8pt}
\caption{External Feature Collection Process and Datasets}
\label{fig:ftr}
\vspace*{-15pt}
\end{figure}

We refer to features that require queries to remote servers as external features. For each URL, we acquire external features by querying two different remote servers:

{\flushleft \bf WHOIS.} We query the WHOIS server responsible for the top level domain of the URL for its registration information, which includes the primary domain name, the registrar, the registrant, and the registration date. We implement our query engine by adopting the \texttt{pywhois} module \cite{pywhois}. Intuitively, the features that come from the WHOIS answers could play an important role in classification. For example, a newly registered site is more likely to be a phishing site as opposed to an old site.

{\flushleft \bf Team Cymru.} We also query Team Cymru server \cite{Cymru} to obtain the network information and the geolocation of each URL. In particular, we obtain the network BGP prefix, the AS number, and the country code. These information are complementary to the former WHOIS information and could potentially help with the classification as well. For instance, multiple phishing URLs are often hosted on the same (badly administered) subnet; as such, the network BGP prefix will give us the desired feature to correlate these sites.

Fig. \ref{fig:ftr} illustrates our external feature collection process.
We note that collecting these external features incur significant latency. On average, the time it takes to collect all external features of an URL in the PhishTank dataset is 1.64 second. The latency depends on a variety of elements, such as, the load of the WHOIS and Team Cymru servers, as well as the geolocations of the WHOIS servers.

%-------------------------- Feature Extraction
\subsection{Feature Extraction}
We now describe our process of extracting lexical and external features and how we prepare them for classification.

\begin{table}[t]
\vspace*{-5pt}
\centering
\caption{Commonly Used URL Obfuscation Techniques from \cite{GareraWorm}}
\vspace*{-8pt}
{\scriptsize
\begin{tabular}{|c|l|}
\hline
{\bf Type} & {\bf Descriptive Examples}\\
\hline
\multirow{2}{*}{I} & {\scriptsize http://210.80.154.30/\~{}test3/.signin.ebay.com/ebayisapidllsignin.html}\\
& {\scriptsize http://0xd3.0xe9.0x27.0x91:3030/.www.paypal.com/uk/login.html}\\
\hline
\multirow{2}{*}{II} & {\scriptsize http://21photo.cn/https://cgi3.ca.ebay.com/eBayISAPI.dllSignIn.php}\\
&  {\scriptsize http://2-mad.com/hsbc.co.uk/index.html}\\
\hline
\multirow{2}{*}{III} &  {\scriptsize http://www.volksbank.de.custsupportref1007.dllconf.info/r1/vm}\\
&  {\scriptsize http://sparkasse.de.redirector.webservices.aktuell.lasord.info}\\
\hline
\multirow{2}{*}{IV} &  {\scriptsize http://www.wamuweb.com/IdentityManagement/}\\
&  {\scriptsize http://mujweb.cz/Cestovani/iom3/SignIn.html?r=7785}\\
\hline
\end{tabular}
}
\label{tab:obfus}
\vspace*{-5pt}
\end{table}

\begin{table}[t]
%\vspace*{-5pt}
\centering
\caption{Lexical Features of a Phishing URL}
\vspace*{-8pt}
{\scriptsize
\begin{tabular}{| l | l | p{4cm} |}
\hline
{\bf URL} & \multicolumn{2}{|l|}{{\color{red} www}.{\color{red} naturenilai}.{\color{blue} com}/{\color{green} form2}/{\color{green} paypal}/{\color{green} webscr}.{\color{BurntOrange} php}?{\color{VioletRed} cmd}=\_{\color{VioletRed} login}}\\
\hline
\multirow{2}{1.4cm}{\bf Auto-Selected} & \multicolumn{2}{| l |}{name=www, name=naturenilai, tld=com, dir=form2, dir=paypal}\\
& \multicolumn{2}{| l |}{file=webscr, ext=php, arg=cmd, arg=login}\\
\hline
\multirow{5}{1.4cm}{\bf Obfuscation-Resistant} & URL & len=54, n\_dot=3, blacklist=1\\
\cline{2-3}
& Domain Name & len=19, IP=0, port=0, n\_token=3, n\_hyphen=0, max\_len=11\\
\cline{2-3}
& Directory & len=14, n\_subdir=2, max\_len=6, max\_dot=0, max\_delim=0\\
\cline{2-3}
& File Name & len=10, n\_dot=1, n\_delim=0\\
\cline{2-3}
& Argument & len=11, n\_var=1, max\_len=6, max\_delim=1\\
\hline
\end{tabular}
}
\label{tab:ftr}
\vspace*{-15pt}
\end{table}

\begin{table*}[t]
\vspace*{-5pt}
\centering
\caption{Summary of Datasets}
\vspace*{-8pt}
{\scriptsize
\begin{tabular}{| l | c | c | c | c | c |}
\hline
{\bf Pairs} & {\bf Yahoo-Phish} & {\bf Yahoo-Malware} & {\bf DMOZ-Phish} & {\bf DMOZ-Malware} & {\bf All Good - All Bad}\\
\hline
{\bf \# Malicious URLs} & 4,082 & 2,001 & 4,082 & 2,001 & 6,083\\
\hline
{\bf \# Legitimate URLs} & 4,143 & 4,143 & 4,012 & 4,012 & 8,155\\
\hline
{\bf \# Lexical Features} & 13,821 & 8,791 & 14,165 & 9,129 & 22,100\\
\hline
{\bf \# External Features} & 18,786 & 16,665 & 9,751 & 7,548 & 24,843\\
\hline
\end{tabular}
}
\label{tab:stat}
\vspace*{-15pt}
\end{table*}

\subsubsection{Lexical Features}
Recall that lexical features can be directly extracted from the URL string. We adopt the approach by Ma {\em et al.} \cite{MaKDD, MaICML} to automatically select binary lexical features. In addition, motivated by the work by Garera {\em et al.} \cite{GareraWorm}, we propose a number of obfuscation-resistant lexical features. We show through empirical results that these features complement the former set of features and help to capture additional obfuscated phishing URLs.

{\flushleft \bf Automatically Selected Features.} The URL string is broken down into multiple tokens. Each token constitutes a binary feature. The delimiters to obtain the tokens are  `/', `?', `.', `=', `\_', `\&', and `-'. Similar to \cite{MaKDD, MaICML}, we distinguish tokens that appear in the domain name, the top level domain, the directory, and the file extension. Different from \cite{MaKDD, MaICML}, we also distinguish tokens that appear in the argument part of the URL. In other words, the same token appearing in different parts of the URL will constitute different binary features. This representation of the URL is known as ``bag-of-word.'' 

{\flushleft \bf Hand-Selected (Obfuscation-Resistant) Features.} In \cite{GareraWorm}, Garera {\em et al.} describe four different URL obfuscation techniques that are commonly used by the attackers: (I) Obfuscating the host with an IP address, (II) Obfuscating the host with another domain, (III) Obfuscating with large host names, and (IV) Domain unknown or misspelled. Table \ref{tab:obfus} illustrates these techniques. Here we propose the following hand-selected lexical features to detect the identified obfuscation techniques; our proposed features are classified into five categories:

{\flushleft \em (i) Features related to the full URL.} These features include the length of the URL, the number of dots in the URL, and whether a blacklisted word appears in the URL. The blacklist we use is similar to the one in \cite{GareraWorm}, which includes the words: {\tt confirm}, {\tt account}, {\tt banking}, {\tt secure}, {\tt ebayisapi}, {\tt webscr}, {\tt login}, and {\tt signin}; and we add the words {\tt paypal}, {\tt free}, {\tt lucky}, and {\tt bonus}. The first two features address Type II obfuscation while the blacklisted words enhance the detection of Type IV obfuscation. 

%\item  
{\flushleft \em (ii) Features related to the domain name.} These features include the length of the domain name, whether an IP address or a port number is used in the domain name, the number of tokens of the domain name, the number of hyphens used in the domain name, and the length of the longest token. %\hspace*{2.7pt} 
These features address obfuscation Type I, Type III, and a technique related to Type III, where hyphens are used instead of dots.

%\item 
{\flushleft \em (iii) Features related to the directory.} These features include the length of the directory, the number of sub-directory tokens, the length of the longest sub-directory token, and the maximum number of dots and other delimiters used in a sub-directory token. These features mainly address Type II obfuscation, where the obfuscated host name is put in the directory. We also observe cases where instead of using `.', the attacker use another character, such as, the underscore, `\_', or dash, `-', as the delimiter of the obfuscated host name. The features in this category address these instances as well.

%\item 
{\flushleft \em (iv) Features related to the file name (page name).} These features include the length of the file name, and the number of dots and other delimiters (`\_' and `-') used in the file name. %\hspace*{2.7pt}  
These features also address Type II obfuscation, but in this case, the obfuscated host name is put in the file name.

%\item 
{\flushleft \em (v) Features related to the argument part.} URLs that serve pages written in server side scripting languages, such as, {\tt php} and {\tt asp}, often have arguments. The features in this category include the length of the argument part, the number of variables, the length of the longest variable value, and the maximum number of delimiters (`.', `\_', and `-') used in a value. %\hspace*{2.7pt} 
We observe that phishing URLs often include a long list of arguments, as well as auto-generated argument values, which are often unusually long. Also, there are instances where the host name is obfuscated in the values assigned to variables. The features here are designed to address these instances.
%\end{enumerate}

Table \ref{tab:ftr} illustrates how we obtain all lexical features.

\subsubsection{External Features}

We extract a number of binary features and one real value feature from the responses we receive from the WHOIS and Team Cymru servers. The registration date gives the real value feature indicating the number of days the site has been up. The other pieces of information that we described in Section \ref{subsec:external} give the binary features.

Finally, for all the real value features, we shift and scale them so that their values lie between 0 and 1. The reason is that we do not want to give a prior preference to any particular feature. We want the weights of the features to be adjusted by the learning algorithms themselves.

\subsection{Summary of Datasets}
We prepare the data for the classification algorithms by combining the legitimate with the malicious URL datasets. In total, we have 5 pairs: Yahoo and PhishTank (Yahoo-Phish); Yahoo and MalwarePatrol (Yahoo-Malware); Open Directory and PhishTank (DMOZ-Phish); Open Directory and MalwarePatrol (DMOZ-Malware); and all good and all bad URLs (All Good - All Bad), where we combine Yahoo with Open Directory and PhishTank with MalwarePatrol. When combining a legitimate dataset with a malicious dataset, we interleave the URLs of the two sets so that the classification algorithms would get a balanced number of instances of both classes when training their models. Table \ref{tab:stat} provides the statistics of these five pairs.

%--------------- Classification Algorithms -----------------------------------------------
\section{Classification Algorithms}
\label{sec:classification}

In this section, we describe four state-of-the-art classification algorithms that we investigate in this work. These include both batch-learning (Support Vector Machine (SVM)) and online learning algorithms (Online Perceptron (OP), Confidence-Weighted (CW), and Adaptive Regularization of Weights (AROW)). All these algorithms come from the machine learning community; to the best of our knowledge, AROW, which turns out to outperform the rest and become our choice, has not been used in the phishing context before. We start by introducing the notation and describing the general difference between batch-based and online classification.

{\flushleft \bf Notation.} Denote the features of an URL as a vector $\vct{x}$ and its label as $y \in \{1,-1\}$, where 1 indicates the URL is malicious and -1 indicates otherwise. A classification algorithm receives a number of data vectors, $\vct{x}_i$, together with their labels, $y_i$, and trains its model based on these labeled data. Then, given a new data vector, $\vct{x}$, the goal of the algorithm is to predict the label, $y$, of this new data based on its trained model. For SVM and OP, the models are a weight vector, $\vct{w}$. For CW and AROW, in addition to $\vct{w}$, the model also includes the covariance matrix of $\vct{w}$, $\Sigma$. For all algorithms, the prediction, $h(\vct{x})$, is the sign of the inner product between $\vct{w}$ and $\vct{x}$:
\begin{align}
h(\vct{x}) = \text{sign}(\vct{w} \cdot \vct{x})
\end{align}

{\flushleft \bf Batched-based vs. Online.} A batch-based algorithm initially trains its model based on a batch of labeled data. It then uses the trained model to predict a number of new data. After some time, it retrains its model based on a new batch of labeled data. Meanwhile, an online classification algorithm continuously retrains its model upon receiving each labeled data and makes prediction of a new data using the latest updated model. Because training a model of a batch-based algorithm requires a batch of data, batch-based algorithms require significantly more memory than online algorithms.

%----------------------------------------------------- SVM
\subsection{Batch Learning}
\subsubsection{Support Vector Machine (SVM)}
The SVMs are widely known for achieving accurate classification of high-dimensional data. They are also shown recently to perform well in the arena of classifying malicious URLs \cite{MaKDD,MaICML}. An SVM constructs a hyperplane that gives the largest distance to the nearest training data points of any class. Finding this hyperplane involves solving an instance of quadratic programming. The label of a new data point is predicted by determining on which side of the hyperplane this point lies.  For a tutorial on SVMs, we refer the reader to \cite{SVMtutor}. In this work, we investigate the performance of batch-based SVMs.

\subsection{Online Learning}

The online algorithms discussed below operate in rounds. In round $t$, an online algorithm receives $\vct{x}_t$ and predicts $\vct{x}_t$'s label as ${\hat y}_t$ using the current model; it then receives the true label, $y_t$, and updates its model based on $(\vct{x}_t, y_t)$.

\subsubsection{Online Perceptron (OP)}
OP updates $\vct{w}$ continuously on error. In particular, $\vct{w}$ is updated if the predicted label, $\hat{y}_t = \text{sign}(\vct{w}_t \cdot \vct{x}_t)$, disagrees with the true label, $y_t$, of $\vct{x}_t$. The update is as follows:
\begin{equation}
\vct{w}_{t+1} \leftarrow \vct{w}_t + y_t\,\vct{x}_t\,.
\end{equation}

OP suffers from a significant drawback: the update rate is fixed and does not take into account the magnitude of classification error. As a result, when making error on prediction, the model may not adapt fast enough to the change of the data,
or it may make a drastic change even when the error is small. Both cases lead to poor classification accuracy.

\subsubsection{Confidence Weighted (CW)}
CW is a linear binary classification algorithm recently introduced by Dredze {\em et al.} \cite{CW}. CW captures the notion of confidence in the weight of a feature. Intuitively, if the weight of a feature does not change very much over time, then one should be more confident that this weight is what it should be. With this confidence notion, CW addresses the drawback of OP through two mechanisms:

First, CW updates the weights of the more confident features less aggressively. For instance, using an IP address in the domain name is a strong indicator of maliciousness; as a result, it does not get updated abruptly over a period of time, thereby having a high confidence value. Then, CW makes sure that this weight will not change much even when it sees an instance of legitimate URL using an IP address in its domain name.

Second, CW does not change the weights too much but just enough to correct for the mistake. In other words, CW updates its model just enough to adapt to the change of the data, while trying to avoid changing too much. The rationale is that the previous model carries a lot of valuable information about the data and should not be changed too abruptly. 

Formally, CW maintains a Gaussian distribution over the weights with mean $\boldsymbol \mu$ and covariance matrix $\Sigma$. The value $\mu_i$ represents what is known about the weight $w_i$, and the value $\Sigma_{i,i}$ captures the confidence in the weight of feature $i$. To classify a new data $\vct{x}$, the weight $\vct{w}$ is drawn from ${\mathcal N} ({\boldsymbol \mu},\Sigma)$. In practice, one can pick $\vct{w} = \boldsymbol \mu$, the average weight vector. The prediction is then as usual: $h(\vct{x}) = \text{sign}(\vct{w} \cdot \vct{x})$.

Unlike OP, CW updates its model, {\em i.e.}, $\boldsymbol \mu$ and $\Sigma$, continuously on every labeled data instead of only when making mistake. This is because making correct prediction also suggests that one should increase his or her confidence of the current weights. The update rule is as follows:
\begin{align}
({\boldsymbol \mu}_{t+1}, \Sigma_{t+1}) = &\text{ arg } \underset{\boldsymbol \mu, \Sigma}{\text{min}} \text{ D}_\text{KL} (\mathcal N ({\boldsymbol \mu}, \Sigma)||\mathcal N ({\boldsymbol \mu}_{t}, \Sigma_{t})) \,,\label{eq:CW-obj} \\
~&\text{ s.t. } \text{Pr}_{\vct{w} \sim \mathcal N (\boldsymbol \mu, \Sigma) } [y_t (\vct{w} \cdot \vct{x}_t)] \ge \eta\,. \label{eq:CW-cstr}
\end{align}

Eq. (\ref{eq:CW-obj}) expresses that the new distribution given by the new $\boldsymbol \mu$ and $\Sigma$ should be as close to the old distribution as possible. The distance between the two distributions is measured by the KL divergence ($\text{D}_\text{KL}$). Eq. (\ref{eq:CW-cstr}) expresses that the update should be enough such that the probability of making correct prediction when seeing the same data in the next round must be bigger than $\eta$, where $\eta$ is a configurable parameter and must be larger than 50\%.

We refer the reader to \cite{CW} for more details. The computational complexity of the update is linear in the number of non-zero features in $\vct{x}_t$. The memory required is constant in the input data, \ie the memory for the current $\vct{x}$.

\subsubsection{Adaptive Regularization of Weights (AROW)}
The final algorithm in this category that we examine is the AROW algorithm by Crammer {\em et al.} \cite{AROW}. AROW can be considered as a modification of CW so that the classifier is more robust in the presence of label noise. For example, if `{\tt whitehouse.gov}' is wrongly labeled as malicious (by an adversary) and fed to CW, then CW will make changes to all features that this URL has so that in the next time slot, if it sees this URL again, it will be likely to flag this URL as malicious. CW, therefore, may drastically increase the weight of the feature ``top level domain is {\tt .gov}''. AROW avoids this drastic behavior by softening the formulation of CW.

Formally, Crammer {\em et al.} \cite{AROW} recast the constraint (\ref{eq:CW-cstr}) of CW as regularizers. The update rule is now as follows:
\begin{align}
({\boldsymbol \mu}_{t+1}, \Sigma_{t+1}) =& \text{ arg } \underset{\boldsymbol \mu, \Sigma}{\text{min}} \text{ D}_\text{KL} (\mathcal N ({\boldsymbol \mu}, \Sigma)\,||\,\mathcal N ({\boldsymbol \mu}_{t}, \Sigma_{t})) \notag \\
~ & + \lambda_1 l_{h^2}(y_t, \boldsymbol \mu \cdot \vct{x}_t) + \lambda_2 \vct{x}_t^T \Sigma \vct{x}_t\,, \label{eq:AROW}
\end{align}
where $l_{h^2}(y_t, \boldsymbol \mu \cdot \vct{x}_t) = (\text{max}\{0,1-y_y(\boldsymbol \mu \cdot \vct{x}_t)\})^2$ is the squared-hinge loss suffered using $\boldsymbol \mu$ to predict the label for $\vct{x}_t$ when its true label is $y_t$, and $\lambda_1$ and $\lambda_2$ are configurable parameters.

Compared to CW, the optimization problem becomes unconstrained. Consider the right hand side of (\ref{eq:AROW}): its first term expresses that the new distribution should be as close to the old distribution as possible. Intuitively, AROW tries to preserve the valuable information of the old model as much as possible. The second term expresses that the new parameters should be able to predict the current example with low loss. Through this term, AROW adapts to the change of the data. Finally, the last term expresses that the confidence in the weights should generally grow.

Similarly to CW, the running time of the update is linear in the number of non-zero features in $\vct{x}_t$. The memory requirement is constant in terms of the input data. We refer the reader to \cite{AROW} for more details. To the best of our knowledge, this is the first time that AROW is used in the phishing context.

%--------------- Evaluation Results ---------------------------------------------
\section{Evaluation Results}
\label{sec:evaluation}

\begin{table}[t]
\vspace*{-5pt}
\centering
\caption{Summary of the Experiments}
\vspace*{-8pt}
{\scriptsize
\begin{tabular}{|c|p{3cm}|p{1cm}|p{1cm}| p{1.5cm}|}
\hline
{\bf \#} & {\bf Description} & {\bf Algorithms} & {\bf Features} & {\bf Datasets}\\
\hline
1 & Compare batch-based algorithms to online algorithms when using lexical features & SVM and AROW & Lexical & Yahoo-Phish\\
\hline
2 & Compare using lexical features to using full features & OP, CW, AROW & Lexical and Full & All pairs\\
\hline
3 & Evaluate the effectiveness of obfuscation-resistant lexical features & AROW & Lexical & All pairs\\
\hline
4 & Evaluate the resilience of AROW to noisy data & CW and AROW & Lexical & Yahoo-Phish with noise\\
\hline
\end{tabular}
}
\label{tab:exp}
\vspace*{-10pt}
\end{table}

We conduct four sets of experiments on various datasets in order to (i) compare batch-based to online algorithms when using just lexical features, (ii) compare using lexical features to using full features, (iii) evaluate the effectiveness of obfuscation-resistant lexical features, and (iv) evaluate the resilience of AROW when working with noisy data. Table \ref{tab:exp} summarizes the experiment scenarios.

\subsection{Batch-Based vs. Online Algorithms}

\begin{figure}[t]
\centering
\includegraphics[width=8cm]{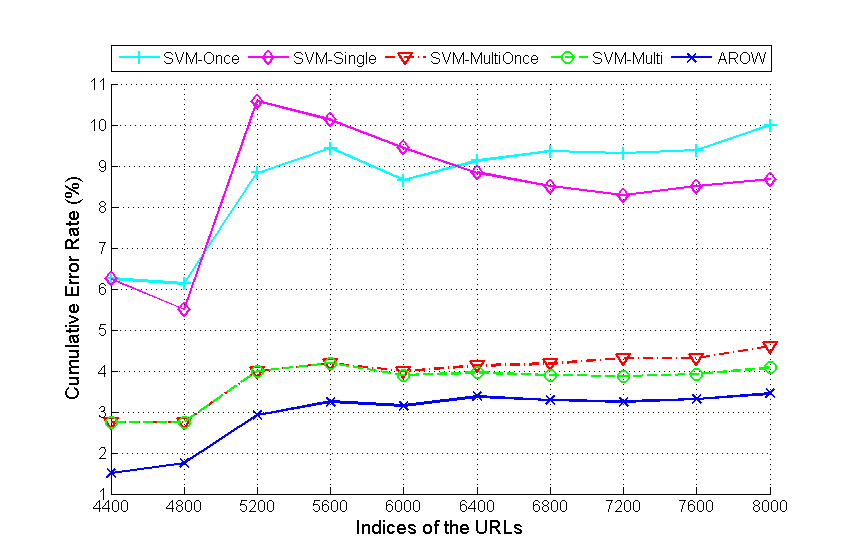}
\vspace*{-10pt}
\caption{Performance of the SVM and the CW Algorithms when using Lexical Features on Yahoo-Phish}
\label{fig:batchVSonline}
\vspace*{-15pt}
\end{figure}

\begin{table}[t]
\vspace*{-5pt}
\centering
{\scriptsize
\caption{Cumulative Error Rate of SVM and CW Algorithms on Yahoo-Phish after the Last URL}
\vspace*{-8pt}
\begin{tabular}{|c|c|c|c|c|}
\hline
\multicolumn{5}{|c|}{\bf Cumulative Error Rate (\%)}\\
\hline
{\bf SVM-Once} & {\bf SVM-Daily} & {\bf SVM-MultiOnce} & {\bf SVM-Multi} & {\bf AROW}\\
\hline
10.00 & 8.68 & 4.60 & 4.08 & 3.45\\
\hline
\end{tabular}
\label{tab:batchVSonline}
}
\vspace*{-10pt}
\end{table}

Here we compare the performance of the batch-based SVM algorithms with the online AROW algorithm when using only lexical features. We present the results of SVMs when using the linear kernel. We use the implementation of linear SVMs by LIBLINEAR \cite{Liblinear}. We implement the AROW algorithm using Matlab based on the closed form update rules in \cite{AROW}. We configure the box constraint $C$ of SVM and the $\lambda$'s of AROW using cross validation on the initialization data. For this set of experiments, $C$ is set to $2^5$, and $\lambda_1$=$\lambda_2$=0.5. For the interest of space, we only present the results of the classification on the Yahoo-Phish dataset (the other datasets give similar results.)

We divide the set of URLs into batches of size 400. For SVM, we examine the performance of the following variants: SVM-Once, SVM-Single, SVM-MultiOnce, and SVM-Multi. For SVM-Once, the model is trained only once on the first batch of URLs. For SVM-Single, the model is retrained after every batch of URLs; however, only one batch is used for every retraining. SVM-MultiOnce is similar to SVM-Once and SVM-Multi is similar to SVM-Single, however, with the size of the batch for training and retraining 10 instead of 1.

The initialization set includes 10 batches, equal to the size of a training batch of SVM-Multi. SVM-Multi, SVM-MultiOnce, and AROW initialize their models using all the URLs in this initialization set. SVM-Once and SVM-Single initialize their models using the last batch of URLs of the initialization set.

Fig. \ref{fig:batchVSonline} shows the cumulative error of the SVM variants and of AROW over time, {\em i.e.}, the ratio of the number of misclassification over the number of classified URLs so far. Table \ref{tab:batchVSonline} gives the cumulative error of these algorithms after the last URL. Based on the plot and the table, we make the following observations.
First, updating the classification models over time is essential as shown by SVM-Single and SVM-Multi outperforming SVM-Once and SVM-MultiOnce, respectively.
Second, training on more data improves the performance, as illustrated by SVM-Multi outperforming SVM-Single. However, we note that there is a fundamental limit on how much data an SVM could train on because of the memory requirement. Moreover, we observe that increasing the training size over 10 batches does not further improve the performance. Third, AROW outperforms all the SVM variants. We believe that this is because AROW is able to adapt to the changes of the URLs quickly as well as retain information of all the features with high confidence values.

{\flushleft \bf Summary.} This set of experiments illustrates that (i) updating the models continuously is essential, and (ii) AROW outperforms SVMs when using lexical features. This superior performance and the light-weight properties (light memory requirement and light computation overhead) make AROW, in particular, and online algorithms, in general, better candidates for the URL classification task. Therefore, in the rest of this section, we focus on the performance of the online algorithms.

\subsection{Lexical Features vs. Full Features}

\begin{figure}[t]
\centering
\includegraphics[width=8cm]{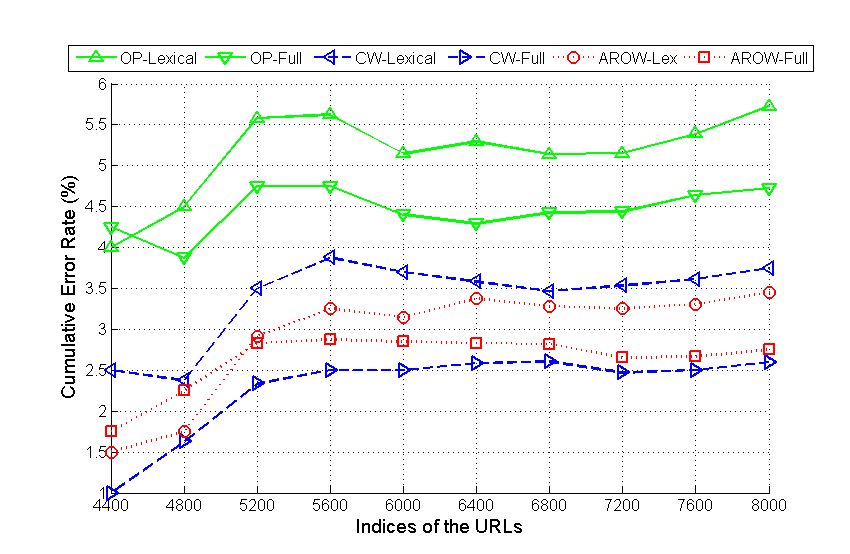}
\vspace*{-10pt}
\caption{Performance of OP, CW, and AROW when using Lexical Features versus when using Full Features on Yahoo-Phish}
\label{fig:lexVSfull}
\vspace*{-15pt}
\end{figure}

\begin{table*}[t]
\vspace*{-5pt}
\centering
{\scriptsize
\caption{Cumulative Error Rate of the OP, CW, and AROW Algorithms on All Datasets after the Last URL}
\vspace*{-8pt}
\begin{tabular}{|l|c|c|c||c|c|c||c|c|c|}
\hline
~ & \multicolumn{3}{c||}{\bf Cumulative Error Rate of OP (\%)} & \multicolumn{3}{c||}{\bf Cumulative Error Rate of CW (\%)} & \multicolumn{3}{c|}{\bf Cumulative Error Rate of AROW (\%)}\\
\cline{2-10}
{\bf Dataset} & {\bf Lexical Ftrs} & {\bf Full Ftrs} & {\bf Gain} & {\bf Lexical Ftrs} & {\bf Full Ftrs} & {\bf Gain} & {\bf Lexical Ftrs} & {\bf Full Ftrs} & {\bf Gain}\\
\hline
Yahoo-Phish & 5.72 & 4.73 & 0.99 & 3.75 ($\eta=73\%$) & 2.60 ($\eta=84\%$) & 1.15 & 3.45 ($\lambda=0.5$) & 2.75 ($\lambda=0.5$) & 0.70 \\
Yahoo-Malware & 3.27 & 2.40 & 0.87 & 2.17 ($\eta=66\%$) & 1.62 ($\eta=66\%$) & 0.55 & 3.05 ($\lambda=5.0$) & 1.98 ($\lambda=5.0$) & 1.07\\
DMOZ-Phish & 6.90 & 4.73 & 2.17 & 3.52 ($\eta=86\%$) & 2.52 ($\eta=79\%$) & 1.00 & 3.60 ($\lambda=50$) & 2.77 ($\lambda=50$) & 0.83\\
DMOZ-Malware & 3.75 & 3.20 & 0.55 & 3.23 ($\eta=58\%$) & 2.35 ($\eta=62\%$) & 0.87 & 3.75 ($\lambda=0.5$) & 2.75 ($\lambda=5.0$) & 1.00\\
\hline
All Good - All Bad & 5.43 & 4.31 & 1.12 & 4.07 ($\eta=58\%$) & 3.14 ($\eta=58\%$) & 0.87 & 5.48 ($\lambda=5.0$) & 4.00 ($\lambda=5.0$) & 1.48\\
\hline
\end{tabular}
\label{tab:lexVSfull}
}
\vspace*{-15pt}
\end{table*}

We conduct the second set of experiments to evaluate how well online classification algorithms do when using only lexical features as opposed to using full features. We examine the performance of the OP, CW, and AROW algorithms on all pairs of datasets. Similar to AROW, we implement OP and CW using Matlab. For the pairs Yahoo-Phish, DMOZ-Phish, and All Good-All Bad, the initialization set includes the first 4000 URLs. For the pairs Yahoo-Malware and DMOZ-Malware, the initialization set is smaller: first 2000 URLs. This is because the total number of URLs in either of these two pairs is smaller than those of the other pairs: 6000 as opposed to over 8000.

Fig. \ref{fig:lexVSfull} plots the cumulative error rates of the algorithms over time for the Yahoo-Phish pair. We omit the plots for the other pairs due to lack of space. Instead, in Table \ref{tab:lexVSfull}, we report the cumulative error rates after the last URL for all pairs. Table \ref{tab:lexVSfull} also reports the configured parameter $\eta$ of each experiment involving CW and $\lambda$ (=$\lambda_1$=$\lambda_2$) for each experiment involving AROW. We note that similar to $\lambda$, $\eta$ is configured using cross validation. Based on the plot and the table, we make the following observations.

 Consider CW and AROW, the cumulative error rates for the pair All Good - All Bad are always larger than all other pairs regardless of using lexical or full features. This suggests that one should build two separate classifiers for PhishTank and MalwarePatrol instead of building a single classifier for both. This agrees with the discussion in Section \ref{sec:dataset}: these are datasets with different characteristics due to their different purposes; therefore, we subsequently focus our discussion on the other four pairs.
For these pairs, CW and AROW outperform OP regardless of using lexical or full features; moreover, CW and AROW have comparable performance when using lexical features: AROW slightly edges CW on Yahoo-Phish while CW slightly edges AROW on the other. Finally, the gain of using full features over lexical features is only about 1\% for both CW and AROW across all the pairs of interest. Using lexical features alone, AROW achieves 96--97\% of accuracy while CW achieves 96--98\% of accuracy.

{\flushleft \bf Summary.} This set of experiments shows that using lexical features alone leads to comparable classification accuracy to full features (only 1\% difference). The high accuracy and the lightweight properties of lexical features make a strong case for using lexical feature alone for the URL classification task.

\subsection{Obfuscation-Resistant (OR) Lexical Features}

\begin{table*}[t]
\vspace*{-5pt}
\centering
{\scriptsize
\caption{Performance of AROW when not using and when using Obfuscation-Resistant (OR) Lexical Features after the Last URL}
\vspace*{-8pt}
\begin{tabular}{|l|c|c|c||c|c|c||c|c|c|}
\hline
~ & \multicolumn{3}{c||}{\bf Cumulative Error Rate (\%)} & \multicolumn{3}{c||}{\bf \# Mis-Classified Benign URLs (FP)}  & \multicolumn{3}{c|}{\bf \# Mis-Classified Malicious URLs (FN)}\\
\cline{2-10}
{\bf Dataset} & {\bf w/o OR Ftrs} & {\bf with OR Ftrs} &  {\bf Gain (Gain Pctg)} & {\bf w/o OR Ftrs} & {\bf with OR Ftrs} &  {\bf Gain} & {\bf w/o OR Ftrs} & {\bf with OR Ftrs} &  {\bf Gain} \\
\hline
Yahoo-Phish & 3.92 ($\lambda=0.5$) & 3.45 ($\lambda=0.5$) & 0.37 (9\%) & 47 & 55 & -8 & 110 & 83 & 27 \\
Yahoo-Malware & 3.70 ($\lambda=5.0$) & 3.05 ($\lambda=5.0$) & 0.65 (18\%) & 88 & 77 & 11 & 60 & 45 & 15 \\
DMOZ-Phish & 4.05 ($\lambda=5.0$) & 3.60 ($\lambda=50$) & 0.45 (11\%) & 50 & 53 & -3 & 112 & 91 & 21 \\
DMOZ-Malware & 5.12 ($\lambda=0.5$) & 3.75 ($\lambda=0.5$) & 1.37 (27\%) & 22 & 41 & -19 & 183 & 109 & 74 \\
\hline
\end{tabular}
\label{tab:autoVShand}
}
\vspace*{-15pt}
\end{table*}

Here we evaluate the effectiveness of the obfuscation-resistant (OR) lexical features when using AROW. Table \ref{tab:autoVShand} reports the performance of AROW when the OR features are not used (only auto-selected features are used) and when the OR features are used. From the results, we can see that the OR features boost the classification accuracy across all pairs of datasets. The reduction of the cumulative error rate ranges from 9\% (on Yahoo-Phish) up to 27\% (on DMOZ-Malware.)

To better understand what is improved, we look at the changes of both the number of mis-classified malicious URLs, {\em i.e.}, false negatives (FNs), and the number of mis-classified benign URLs, {\em i.e.}, false positives (FPs). From Table \ref{tab:autoVShand}, we can see that the improvement mainly comes from the reduction of the number of mis-classified malicious URLs. In particular, we reduce the number of FNs (ranging from 15 to 74) for a modest increase in the number of FPs (ranging from 3 to 19.) In the context of protecting users from phishing URLs, having a few false alarms is arguably better than missing many malicious sites. We note that this is not necessarily the case in other contexts, \eg spam.

{\flushleft \bf Summary.} These experiments show that the obfuscation-resistant lexical features effectively improve the overall classification accuracy by catching more phishing URLs.

\subsection{The Resilience of AROW to Noisy Data}

\begin{figure}[t]
\vspace*{-5pt}
\centering
\includegraphics[width=8cm]{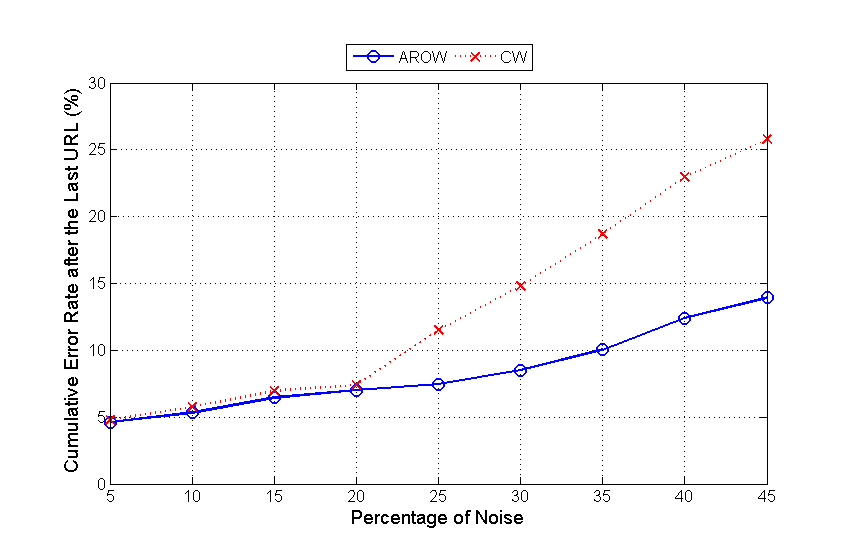}
\vspace*{-10pt}
\caption{Performance of AROW and CW on Yahoo-Phish with Various Amount of Noise}
\label{fig:noise}
\vspace*{-15pt}
\end{figure}

In this last set of experiments, we examine how resilient AROW is to noisy data. Here we report the results on Yahoo-Phish (the other pairs give similar results.) To create noise, we randomly select a number of URLs and change their labels from malicious to benign or vice versa. Fig. \ref{fig:noise} shows the cumulative error after the last URL of both AROW and CW on Yahoo-Phish with various amount of noise. We make the following observations:
First, AROW consistently achieves better classification accuracy than CW; moreover, the noisier the dataset, the larger the difference between the performance of AROW and CW. Second, AROW is able to maintain very high accuracy (about 95\%) when there is a modest amount of noise (from 5 to 10\%) and high accuracy (above 90\%) even when there is a moderate amount of noise (from 10 to 30\%.) 

{\flushleft \bf Summary.} 
AROW can achieve high classification accuracy, higher than CW, when working with noisy data.

%--------------- Understanding the Performance ----------------------------------------------------
\section{Understanding the Performance}
\label{sec:analysis}

In this section, we pose two questions: (i) Why do online algorithms outperform batch-based algorithms in classifying phishing URLs? And, (ii) why do the advanced online algorithms, namely, AROW and CW, outperform the classical OP algorithm in this context? Note that we have already described the theoretical differences among these algorithms in Section \ref{sec:classification}, which may partially answer the questions posed. However, in this section, we seek to better understand how characteristics of phishing URLs affect the performance of these learning algorithms. In particular, we examine the importance of long term memory and the importance of fast model update. The dataset we will use in this section is PhishTank.

{\flushleft \bf Notation.} First, we introduce the notion of {\em similarity} between two URLs. We consider two URLs $u$ and $v$ similar, denoted by $u \sim v$, if their number of common binary lexical features exceeds a threshold $\tau$. For example, for $\tau=3$, the following two URLs are considered similar:\\
\hspace*{3mm} $\bullet$ {\tt 67.23.226.61/$\sim$sarsefil/Absa/index.html}\\
\hspace*{3mm} $\bullet$ {\tt 67.23.226.61/$\sim$sarsefil/index.html}\\
This is because they share at least 4 binary features: an IP is used in the domain name, one directory token is {\tt $\sim$sarsefil}, the file name is {\tt index}, and the file extension is {\tt html}. We focus on binary features because of their interpretability and their dominance in the feature set. 

Subsequently, for each URL $u_i$, where $i$ is the order of $u$, we find all URLs that come before it and are similar to it. We call the number of URLs between the latest URL that is similar to $u_i$ and $u_i$ the {\em minimum distance of similarity}, denoted by $\delta_\text{min}(u_i)$. Similarly, we call the number of URLs between the earliest URL that is similar to $u_i$ and $u_i$ the {\em maximum distance of similarity}, denoted by $\delta_\text{max}(u)$. Formally,
\begin{align}
\delta_\text{min}(u_i) =& \underset{j}{\text{ min }} (i-j) \text{, s.t. } u_i \sim u_j, j<i\\
\delta_\text{max}(u_i) =& \underset{j}{\text{ max }} (i-j) \text{, s.t. } u_i \sim u_j, j<i
\end{align}
If there is no URL similar to $u_i$, its default values for $\delta_\text{min}(u_i)$ and $\delta_\text{max}(u_i)$ are 0 and $n+1$, respectively, where $n$ is the number of URLs in the dataset.

\subsection{The Importance of Long Term Memory}

\begin{figure}[t]
\vspace*{-5pt}
\centering
\includegraphics[width=8cm]{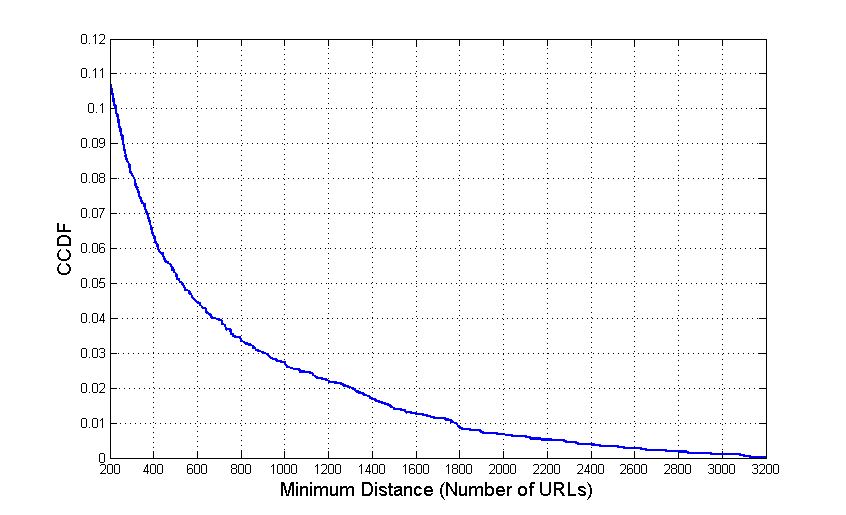}
\vspace*{-10pt}
\caption{The Tail of the Complementary Cumulative Distribution Function (CCDF) of the Minimum Distance of Similarity of All PhishTank URLs}
\label{fig:min}
\vspace*{-15pt}
\end{figure}

Fig. \ref{fig:min} shows the tail of the complementary cumulative distribution function (CCDF) of the minimum distance of similarity of all PhishTank URLs, {\em i.e.}, the CCDF of $\delta_\text{min}(u_i), \forall i$, when $\tau=3$. We observe that there is a significant number of URLs (about 10\%) whose minimum distance of similarity are larger than 200. This means that if the batch size is limited to 400 (200 benign and 200 phishing URLs) then these 10\% of URLs will not have any similar URLs in the batch. This reduces the classification accuracy because the classification model, in this case, has not yet learned about any URL that is similar to these URLs by the time it needs to classify them. Fig. \ref{fig:min} also shows that there is a small number of URLs (under 1\%) whose $\delta_\text{min}$ are above 2000. This explains why increasing the batch size of SVM-Multi above 4000 (2000 benign and 2000 malicious URLs) does not produce any significant improvement.

In general, Fig. \ref{fig:min} demonstrates that phishing URLs require long term memory. This explains why extending the batch size results in higher classification accuracy. When using batch-based algorithms, the size of the batch is, however, limited by the amount of memory available. Online algorithms, on the other hand, do not have this limitation. In fact, they retain information about all of the URLs that they have seen. In this sense, online algorithms effectively have infinite batch size. This explains why they have an edge over batch-based algorithms in this context.

\subsection{The Importance of Fast Model Update}

\begin{figure}[t]
\vspace*{-5pt}
\centering
\includegraphics[width=8cm]{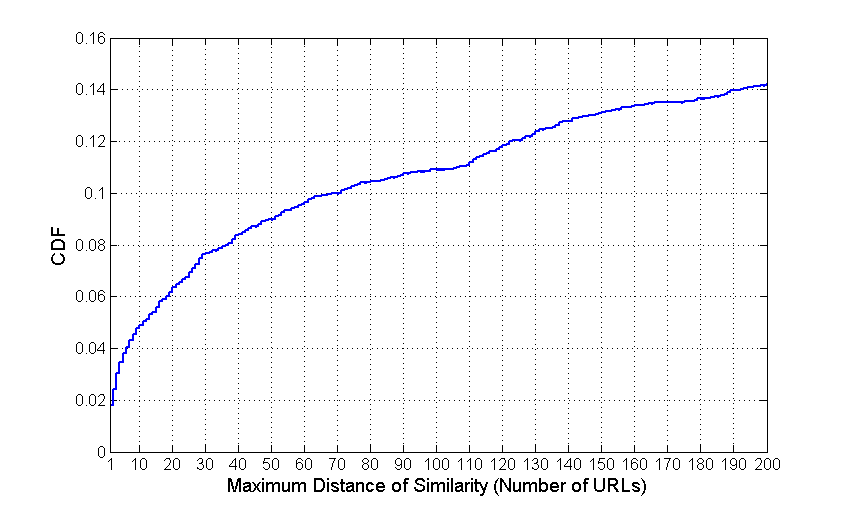}
\vspace*{-10pt}
\caption{The Head of the Cumulative Distribution Function (CDF) of the Maximum Distance of Similarity of All PhishTank URLs}
\label{fig:max}
\vspace*{-15pt}
\end{figure}

Fig. \ref{fig:max} shows the head of the cumulative distribution function (CDF) of the maximum distance of similarity of all PhishTank URLs, {\em i.e.}, the CDF of $\delta_\text{max}(u_i), \forall i$, when $\tau=3$. We observe that there is a significant number of phishing URLs (about 10\%) whose maximum distance of similarity are smaller than 100. This means that for these 10\% of URLs, their only similar URLs are within 100 recent URLs. When using batch-based algorithms with batch size of 400, the model only updates every 200 malicious URLs. This means that the URLs similar to those 10\% of URLs are more than 50\% not likely to be learned by the model by the time it needs to classify those 10\% of URLs. This negatively affect the classification performance. Therefore, it is essential to update the model more often. 

Fig. \ref{fig:max} also shows that about 2\% of URLs having maximum distance of similarity of 1. This indicates two points. First, unless the model is updated after every malicious URL, which is prohibitive expensive for batch-based algorithms, the model will not learn about any URL that is similar to these 2\% of URLs by the time it needs to classify them. This shows why even with infinite amount of memory, SVM algorithms may still fall behind AROW and CW. Second, unless the model is updated rapidly to reflect the recent features present in the last malicious URL, the model will not be able to effectively classify about 2\% of URLs. This demonstrates the edge that AROW and CW have over OP. In fact, the OP algorithm may not reflect the necessary changes within a number of URLs due to its simple update rule.

{\flushleft \bf Summary.} In this section, we discussed why the AROW and CW algorithms outperform the SVM and OP algorithms on phishing datasets. The two main reasons are that AROW and CW can retain the long-term history of all the URLs they have seen, and at the same time, can update their models rapidly to catch new trends of phishing URLs.

%\section{Implementation and Deployment}
\section{PhishDef Deployment}
\label{sec:deployment}

{\flushleft \bf PhishDef.} Based on the results of the evaluation, we propose \phishdef, a system which implements the AROW algorithm and uses only lexical features to classify URLs. By implementing the AROW algorithm, \phishdef~is able to achieve high classification accuracy even when working with noisy data, and at the same time, being lightweight in terms of both computation and memory requirement. By using only lexical features, \phishdef~reduces the page loading latency and avoids reliance on remote servers. \phishdef~is able to perform on-the-fly classification of URLs and protect Internet users from malicious URLs. The high accuracy of \phishdef~has already been demonstrated in the thorough comparisons of classification algorithms and features in Sections \ref{sec:evaluation} and \ref{sec:analysis}. In the rest of this section, we give guidelines for potential deployment.

\begin{figure}[t]
\centering
\includegraphics[width=6cm]{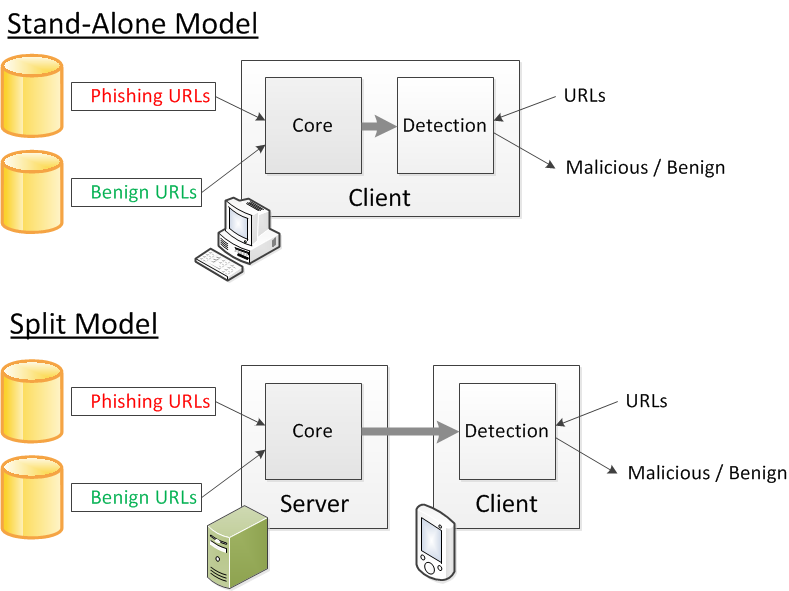}
\vspace*{-5pt}
\caption{PhishDef Deployment Models}
\label{fig:deploy}
\vspace*{-15pt}
\end{figure}

{\flushleft \bf Deployment Options.} \phishdef~can be divided into two main components: the {\em core component}, which maintains and updates the classification model, and the {\em detection component}, which uses the classification model to classify newly input URL. There are two possible deployment models for \phishdef, depicted in Fig. \ref{fig:deploy}, based on where these two components are maintained.

{\flushleft  \em Stand-Alone.} One option is that both components are maintained at the client side. The core component runs as a background service, maintaining and updating its model by querying labeled benign and malicious URLs from, {\em e.g.}, Yahoo Directory and PhishTank, respectively. The detection component runs as a browser add-on, classifying URL on-the-fly using the latest model from the core component.

{\flushleft  \em Split.} Another option is that the core component runs on a separate server, maintaining and updating the classification model as usual. The detection component runs as a browser add-on; however, in this case, before the user browses the Internet, the detection component needs to download the model (the weight vector) and the dictionary of features (to extract features from new URLs) from the server.

The first model has the advantage of being independent, not relying on a separate server. However, it suffers from several drawbacks: (i) the amount of traffic required to update the models, {\em e.g.}, traffic to and from Yahoo Directory and PhishTank, scales linearly with the number of users; and (ii) it is not appropriate for mobile devices, such as smart phones, since it is wasteful of bandwidth and battery life to keep a service running in the background; moreover, mobile devices may not have persistent Internet connection to keep the classification models up-to-date. We believe that the second option is more practical because it saves bandwidth (the size of the weight vector and the dictionary is much smaller than the size of the labeled URLs) and it is appropriate for mobile devices as no background process is needed. While the focus of this paper was on the evaluation of various candidate techniques and feature sets for URL classification, in future work, we will develop and make publicly available add-ons for Firefox and Chrome that implement \phishdef~functionality.

%--------------- Conclusion ---------------------------------------------------
\section{Conclusion}
\label{sec:conclusion}
In this work, we proposed \phishdef~-- a proactive defense scheme which provides the Internet users with protection against phishing attacks. \phishdef~ works by detecting phishing URLs on-the-fly at the client side using only lexical features. \phishdef~is highly accurate (97\%), avoids the overhead of querying remote servers, is lightweight in terms of both computation and memory requirement, and is resilient to noisy training data. 

% -------------- Bibliography --------------------------------------------------
\bibliographystyle{abbrv} % ACM style REMEMBER to copy content from .bbl over to crate a self-contained file
\bibliography{MalURLWork}

\end{document}